\documentclass[
    ,final            % use final for the camera ready runs
  ]
  {aipproc}

\layoutstyle{6x9}

\begin{document}

\title{Simulations of AGN feedback in galaxy clusters and groups}


\author{Ewald Puchwein}{
  address={Max-Planck-Institut f\"{u}r Astrophysik,
      Karl-Schwarzschild-Stra\ss{}e 1, 85740 Garching bei
      M\"{u}nchen, Germany}
}

\author{Debora Sijacki}{
  address={Institute of Astronomy, Madingley
      Road, Cambridge, CB3 0HA, United Kingdom}
}

\author{Volker Springel}{
  address={Max-Planck-Institut f\"{u}r Astrophysik,
      Karl-Schwarzschild-Stra\ss{}e 1, 85740 Garching bei
      M\"{u}nchen, Germany} 
}

\begin{abstract}
There is compelling evidence that black holes (BHs) in cluster centers
vigorously interact with their surroundings,
indicating that any realistic model of cluster formation needs to
account for these processes. Here we use high-resolution cosmological
simulations of a large cluster and group sample to study how BHs affect
their host systems. We focus on two specific properties, the halo gas
fraction and the X--ray luminosity-temperature scaling relation, both of
which are notoriously difficult to reproduce in self-consistent
hydrodynamical simulations. We show that BH feedback can solve both of
these issues, bringing them in excellent agreement with observations,
without alluding to the `cooling only' solution that produces
unphysically bright central galaxies.
\end{abstract}

\maketitle

\input{aas_macros.sty}

%%%%%%%%%%%%%%%%%%%%%%%%%%%%%%%%%%%%%%%%%%%%
%% MAINMATTER
%%%%%%%%%%%%%%%%%%%%%%%%%%%%%%%%%%%%%%%%%%%%

\section{Introduction}

Hydrodynamical cluster simulations which only
include radiative cooling typically suffer from excessive overcooling
within the densest cluster regions. This results in a very
large fraction of cold gas and consequently
a large amount of stars that would form out of it. While the associated
removal of low-entropy gas breaks the
self-similarity of the cluster scaling relations in a way that resembles
observations \citep{Bryan2000, Dave2002, Nagai2007}, this `cooling-only'
scenario is generally discounted because observed clusters do not show
such strong cooling flows and enormously bright central galaxies.
Instead, it is widely believed that some non-gravitational energy input
strongly affects the intracluster medium (ICM).

There has been considerable effort \citep[see][and references
therein]{Borgani2004} to include feedback mechanisms associated with
star formation in hydrodynamical simulations in order to reduce
excessive overcooling in cluster cores. However, so far simulation
models have failed to simultaneously reproduce the masses and colors of
central galaxies, the observed temperature and metallicity profiles, as
well as the observed X--ray luminosity-temperature ($L_{\rm X}-T$)
relation. In particular, the X--ray luminosities have been found to be
substantially larger than the observed values \citep{Borgani2004} on the
group scale.

Feedback from active galactic nuclei (AGN) might resolve the
discrepancies by providing a heating mechanism that offsets cooling,
lowers star formation rates and removes gas from the centers of poor
clusters and groups, thereby reducing their X--ray luminosities to
values compatible with observations. We investigate whether AGN
feedback can indeed solve these problems by performing simulations
of galaxy clusters and groups that employ a state-of-the-art model
\citep{Sijacki2007} for BH growth and feedback.

\section{The simulations}

We have performed high-resolution cosmological hydrodynamical
simulations of a large galaxy cluster and group sample,
accounting for radiative cooling, UV heating, star
formation and supernovae feedback. For each halo, two kinds of
simulations were performed. One containing the physics just described
and an additional one including a model for BH growth and associated
feedback processes as in \cite{Springel2005b} and \cite{Sijacki2007}.
This allows us to compare the very same clusters simulated with and
without AGN heating in order to clearly pin down the imprints of AGN
activity on cluster and group properties.

The simulations are described in detail in \cite{Puchwein2008}. Here we
just summarize the main features of our BH growth and
feedback model. We assume that low-mass seed BHs are produced
sufficiently frequently that any halo above a certain threshold mass
contains one such BH at its center. In the simulations,
an on-the-fly friends-of-friends group finder puts seed BHs with a mass
of $10^5 h^{-1} M_\odot$ into halos when they exceed a mass of $5
\! \times \! 10^{10} h^{-1} M_\odot$ and do not contain any BH yet. The
BHs are then allowed to grow by mergers with other BHs and by accretion
of gas at the Bondi-Hoyle-Lyttleton rate, but with the Eddington limit
additionally imposed.

We use two distinct feedback models depending on the BH accretion rate
(BHAR) \citep[see][]{Sijacki2007}. For large accretion rates above
$0.01$ of the Eddington rate, AGN feedback is assumed to occur in a
``quasar-mode'', where only a small fraction of the luminosity is
thermally coupled to the ICM. We adopt this thermal heating efficiency
to be $0.5\%$ of the rest mass-energy of the accreted gas, which
reproduces the observed BH mass-bulge stellar
velocity dispersion relation \citep{DiMatteo2005}. For lower BHARs,
 we assume that feedback is in a so-called ``radio-mode'', where AGN
jets inflate hot, buoyantly rising bubbles in the surrounding ICM. The
duty cycle of bubble injection, energy content of the bubbles as well as
their initial size are determined from the BHAR. We assume the
mechanical feedback efficiency provided by the bubbles to be $2\%$ of
the accreted rest mass-energy, which is in good agreement with
observations of X--ray luminous elliptical galaxies \citep{Allen2006}.

It was shown in \cite{Sijacki2007} that this model leads to a
self-regulated BH growth and brings BH and stellar mass densities into
broad agreement with observational constraints.

Finally, we obtain realistic X--ray luminosities and spectroscopic
temperatures for each simulated halo, as described
in \cite{Puchwein2008}.

\section{Halo gas fractions}

\begin{figure}
%\begin{center}
 \scalebox{0.71}{\includegraphics{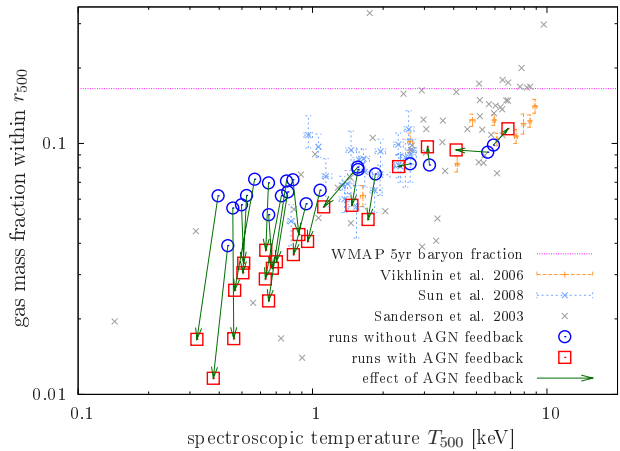}}
%\end{center}
  %\plotone{f1.eps}%  
\caption{Halo gas mass fractions within $r_{500}$ of clusters
    and groups simulated without (circles) and with AGN
feedback (squares). The arrows illustrate the effect of the AGN
heating for each halo. For comparison, we show constraints obtained from
X-ray observations.}
\label{fig:gas_fractions}
\end{figure}

In Fig.~\ref{fig:gas_fractions}, we show the ratio of
gas mass to total mass inside the radius $r_{500}$ as a function of halo
X--ray temperature. 
%\citep{Vikhlinin2006,Sun2008,Sanderson2003}.
The most obvious effect of the AGN feedback is the significantly reduced
gas fraction at the low temperature end of our sample, i.e. in poor
clusters and groups. There the AGN heating drives a substantial fraction
of the gas to radii outside of $r_{500}$. This lowers halo gas fractions
in spite of the reduced fraction of gas that is converted into stars in
the runs with AGN. The potential wells of massive clusters are, on the
other hand, too deep for AGN to efficiently remove gas from
them. Thus the effect of the suppressed star formation becomes more
important towards more massive systems. While the gas
fraction in the very inner regions of massive clusters is somewhat
reduced by the AGN, we find it unchanged or slightly increased within
$r_{500}$.

\section{The $L_{\rm X}-T$ relation}

\begin{figure}
  \scalebox{0.72}{\includegraphics{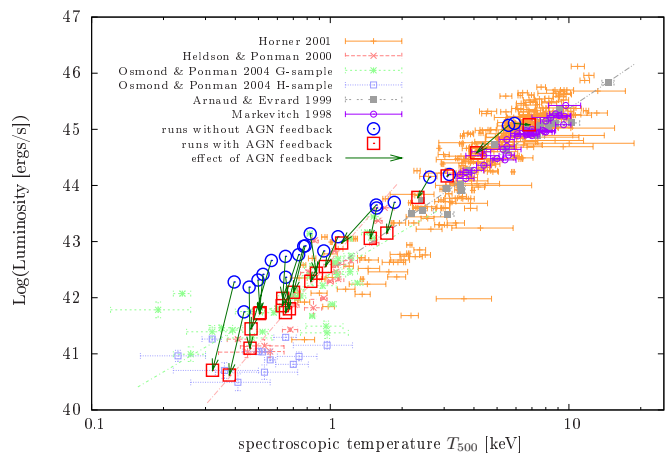}}%
\caption{X--ray luminosity as a function of spectroscopic temperature
  for clusters and groups simulated without (circles)
  and with AGN feedback (squares). For each halo, the arrow illustrates
the effect of the AGN heating. Data from observational X-ray studies is
shown for comparison. For a subset also the best-fit power-law
  $L_{\rm X}-T$ relation is plotted. Including AGN feedback into the
  simulations drastically improves the agreement with observations and
  resolves the discrepancy on the group scale.}
\label{fig:L_X-T_relation}
\end{figure}

By removing gas from the centers of poor clusters and
groups, AGN heating also suppresses their X--ray luminosities and
affects the $L_{\rm X}-T$ relation. In Fig.~\ref{fig:L_X-T_relation}, we
plot the X--ray luminosities $L_{500}$ against the spectroscopic
temperatures $T_{500}$, for all our halos.
%Data from
%observational X--ray studies is shown for comparison \citep{Horner2001,
%Helsdon2000, Osmond2004, Arnaud1999, Markevitch1998}.

Without AGN, we obtain substantially larger X--ray
luminosities for poor clusters than observed, while for massive clusters
there is reasonable agreement. This finding is in line with
previous studies \citep{Borgani2004}, and is a manifestation of the
long-standing problem to explain cluster scaling relations in
hydrodynamical simulations. However, when we include the AGN feedback
the discrepancies between simulated and observed $L_X-T$ relation are
resolved. In particular, the relation is
steepened on the group scale, as AGN heating removes a
larger fraction of gas from smaller halos and thereby reduces their
X--ray luminosities. For massive clusters, the effect of AGN is
less important. Overall, the $L_X-T$ relation obtained from the
simulations with AGN feedback is consistent with observations at all
mass scales, from massive clusters to small groups.

\section{Summary and conclusions}

We have performed very high-resolution simulations of
a galaxy cluster and group sample and compared simulations
with and without AGN feedback. We find that:
\begin{itemize}
\item AGN feedback significantly lowers the gas mass fractions in poor
  clusters and groups, even though fewer baryons are turned into stars.
This is because the AGN heating drives gas from
  halo centers to their outskirts. Both the gas and
  stellar fractions are in a much better agreement with observations
when AGN  are included.
\item AGN feedback significantly reduces the X--ray luminosities of poor
  clusters and groups, while the X--ray temperature stays roughly the
same or is even slightly reduced. This results in a steepening of the
$L_{\rm X}-T$ relation on the group scale.
\item The $L_{\rm X}-T$ relation obtained from simulations with AGN
feedback is in excellent agreement with observations at all mass
scales.
\end{itemize}

We find it extremely encouraging that this simple model for BH growth
and feedback is capable of bringing the analyzed
properties of galaxy clusters and groups into much better agreement with
observations. This resolves a long-standing problem in their
hydrodynamical modeling and opens up new exciting possibilities for
using simulations to investigate their properties and their
co-evolution with central supermassive BHs.

%\begin{figure}
%  \includegraphics[height=.3\textheight]{golfer}
% \caption{Picture to fixed height}
%\end{figure}

\begin{theacknowledgments}
We thank Simon White, Martin Haehnelt,
Klaus Dolag and Gabriel Pratt for constructive discussions. D.S.
acknowledges Postdoctoral Fellowship from the STFC. Part of the
simulations were run on the Cambridge HPC cluster Darwin.
\end{theacknowledgments}



\bibliographystyle{aipprocl}   % if natbib is available

\bibliography{paper}

\end{document}